\documentstyle[epsfig,amssymb,aps,preprint]{revtex}
\tightenlines
\begin{document}
\def\dr{\rangle\!\rangle}
\def\dl{\langle\!\langle}
\def\ad{a^{\dag}}\def\bd{b^{\dag}}\def\hU{\hat U}\def\hUd{\hat U^{\dag}}
\def\<{\langle}\def\>{\rangle}\def\hb{\hbar}\def\cd{c^{\dag}}
\def\hr{\hat\varrho}
\def\d{\mbox{d}}
\def\pni{\par\noindent}
\title{On the correspondence between classical and quantum
measurements on a bosonic field} 
\author{G. M. D'Ariano and
M. F. Sacchi} \address{Theoretical Quantum Optics Group\\ INFM ---
Unit\`a di Pavia \& Dipartimento di Fisica ``A. Volta''\\
Universit\`a degli Studi di Pavia, \\ via A. Bassi 6, I-27100 Pavia,
ITALY} \author{H. P. Yuen}
\address{Department of Electrical
and Computer Engineering, \\ Department of Physics and Astronomy,\\
Northwestern University \\ 2145 North Sheridan Road, Evanston, IL
60208-3118 USA} 
\maketitle
\begin{abstract}
We study the correspondence between classical and quantum measurements
on a harmonic oscillator that describes a one-mode bosonic field
with annihilation and creation operators $a$ and $a^{\dag}$ with
commutation $[a,a^{\dag}]=1$. We connect the quantum measurement of an
observable $\hat O=\hat O(a,a^{\dag})$ of the field with the
possibility of amplifying the observable $\hat O$ ideally through a
quantum amplifier which achieves the Heisenberg-picture evolution
$\hat O\to g\hat O$, where $g$ is the gain of the amplifier.
The ``classical'' measurement of $\hat O$ corresponds to the joint
measurement of the position $\hat q={1\over2}(a^{\dag}+a)$ and
momentum $\hat p={i\over2}(a^{\dag}-a)$ of the harmonic oscillator,
with following evaluation of a function $f(\alpha,\bar \alpha )$ of the
outcome $\alpha=q+ip$. For the electromagnetic field the joint
measurement is achieved by a heterodyne detector. The quantum
measurement of $\hat O$ is obtained by preamplifying the heterodyne
detector through an ideal amplifier of $\hat O$, and rescaling the outcome
by the gain $g$. We give a general criterion which states when this
preamplified heterodyne detection scheme approaches the ideal quantum
measurement of $\hat O$ in the limit of infinite gain. We show
that this criterion is satisfied and the ideal measurement is achieved
for the case of the photon number operator $a^{\dag}a$ and for the
quadrature $\hat X_{\phi}=(\ad\,e^{i\phi}+a\,e^{-i\phi})/2$, where one
measures the functions $f(\alpha,\bar \alpha )=|\alpha|^2$ and
$f(\alpha,\bar \alpha )=\mbox{Re}(\alpha e^{-i\phi})$ of the field,
respectively. For the photon number operator $a^{\dag}a$ the
amplification scheme also achieves the transition from the continuous
spectrum $|\alpha|^2\in\mathbb R$ to the discrete one $n\in\mathbb N$
of the operator $a^{\dag}a$. Moreover, for both operators $a^{\dag}a$
and $\hat X_{\phi}$ the method is robust to nonunit quantum
efficiency of the heterodyne detector.  On the other hand, we show
that the preamplified heterodyne detection scheme does not work for
arbitrary observable of the field. As a counterexample, we prove that
the simple quadratic function of the field $\hat K=i(a^{\dag
2}-a^2)/2$ has no corresponding polynomial function $f(\alpha,\bar 
\alpha )$---including the obvious choice
$f=\hbox{Im}(\alpha^2)$---that allows the measurement of $\hat K$
through the preamplified heterodyne measurement scheme.\\
\small{1999 PACS number(s): 03.65.-w, 03.65.Bz, 42.50.Dv, 42.50-p}
\end{abstract}
\newpage
\section{Introduction}
In the standard formulation of Quantum Mechanics an abstract
concept of physical observable is formulated in terms of real
eigenvalues and sharp probability distributions, which leads to the
well known correspondence between observables and self-adjoint 
operators on the Hilbert space \cite{vn}.  A natural extension of this
formulation is based on the general concept of Positive
Operator-Valued Measure (POVM) \cite{dav,hel}, which allows the
description of joint measurements of non-commuting observables, with
generally complex eigenvalues and probability distributions that are
not sharp for any quantum state. From an operational point of view,
however, we have no prescription on how to achieve the ideal quantum
measurement (i. e. with minimum noise) of a generic operator, and the 
problem of finding a {\em universal detector} is still an open
one. Quantum homodyne tomography---the only known method for 
measuring the state itself of the field---can also be regarded as a
kind of universal detection \cite{tokio}, however it is far
from being ideal, due to the occurrence of statistical measurement
errors that are intrinsic of the method. 
\par In this paper we study the possibility of achieving the ideal
measurement of an observable $\hat O=\hat O(a,a^{\dag})$ of one mode
of the electromagnetic field by means of a fixed detection
scheme---the heterodyne detector---after ideal preamplification $\hat
O\to g\hat O$ of the observable $\hat O$, $g$ denoting the amplifier gain, 
seeking a connection between the problem of measuring $\hat O$ and
that of amplifying $\hat O$ ideally. As heterodyne detection corresponds to 
the ideal joint measurement of the canonical pair $\hat q={1\over2}
(a^{\dag}+a)$ and $\hat p={i\over2}(a^{\dag}-a)$ of a harmonic
oscillator in the phase space, in this way we also try to set a link between
classical and quantum measurements. We will give a necessary and
sufficient condition that establishes when the preamplified heterodyne
detection scheme approaches the ideal quantum measurement of $\hat O$
in the limit of infinite gain. We show that such condition is satisfied for the photon  
number operator $a^{\dag}a$---corresponding to the function
$f(\alpha,\bar \alpha )=|\alpha|^2$ of the heterodyne outcome
$\alpha\in\mathbb C$---and for the quadrature operator $\hat
X_{\phi}=(\ad\,e^{i\phi}+a\,e^{-i\phi})/2$---corresponding to the
function $f(\alpha,\bar \alpha )=\mbox{Re}(\alpha e^{-i\phi})$.
For the photon number operator $a^{\dag}a$ the amplification scheme
also achieves the transition from the continuous spectrum
$|\alpha|^2\in\mathbb R$ to the discrete spectrum 
${\mathbb S}_{a^{\dag}a}\equiv{\mathbb N}$ of $a^{\dag}a$.
Moreover, for both operators $a^{\dag}a$ and $\hat
X_{\phi}$ the methods is also robust to nonunit quantum efficiency of
the heterodyne detector.  On the other hand, we will see that the
preamplified heterodyne scheme does not work for arbitrary
observable of the field. As a counterexample, we show that,
unexpectedly, the simple quadratic function of the field $\hat
K=i(a^{\dag 2}-a^2)/2$ has no corresponding polynomial function
$f(\alpha,\bar\alpha )$---including the obvious choice
$f=\hbox{Im}(\alpha^2)$---which allows the measurement of $\hat K$
through the preamplified heterodyne measurement scheme.
\par The paper is organized as follows. In Section \ref{s:hetero} we
derive the POVM of the heterodyne measurement of a function $f$ of the
field, for generally nonunit quantum efficiency. In Section \ref{s:ideampl} 
we analyze the ideal amplification of an observable $\hat O$, and
prove that it can be always achieved by a unitary transformation. In
Section \ref{s:criter} we give a necessary and sufficient condition 
for the preamplified heterodyne detection scheme to approach the
ideal measurement of $\hat O$. Section \ref{s:examples} is devoted to the two
examples $\hat O=a^{\dag}a$ and $\hat O=\hat X_{\phi}$ which satisfy
the requirements of the general criterion of Sect. 
\ref{s:ideampl}. There we also prove explicitly that the ideal
measurement of $\hat O$ is achieved by the 
preamplified heterodyne detection scheme in the limit of infinite gain
of the amplifier, also for nonunit quantum
efficiency of the heterodyne detector. Section \ref{s:counterex} is
devoted to the counterexample $\hat K=i(a^{\dag 2}-a^2)/2$, where,
in order to prove that the preamplified heterodyne scheme does
not work, we also derive the explicit analytical form of the ideal
amplification map for $\hat K$. Section \ref{s:conclu} concludes the
paper by summarizing the main results.
\section{Heterodyne detection}\label{s:hetero}
\par Heterodyne detection corresponds to measuring the complex field
$\hat Z=a+b^{\dag}$, $a$ and $b$ denoting the signal and the
image-band modes of the detector, respectively. The measurement is an
exact joint measurement of the commuting observables $\hbox{Re}\,\hat Z$
and $\hbox{Im}\,\hat Z$, but can
also be regarded as the joint measurement of the non commuting
operators $\hbox{Re}a$ and $\hbox{Im}a$, by considering the image-band
mode in the vacuum state. In this way the vacuum
fluctuations of $b$ introduce an additional 3dB noise, which can be
proved to be the minimum added noise in an ideal joint measurement of
a conjugated pair of non commuting observables \cite{yuenjoint}.  
\par The probability density in the complex
plane $p(\alpha ,\bar \alpha )$ for heterodyne detection is given by
the Fourier transform of the generating function of the moments of
$\hat Z$, namely
\begin{eqnarray}  
p(\alpha ,\bar \alpha )= \int {\d^2\lambda\over\pi^2}\,\langle
e^{\lambda \hat Z^{\dag}-\bar\lambda \hat Z}\rangle
\,e^{\bar\lambda \alpha -\lambda\bar \alpha } \doteq \langle
\delta^{(2)}(\alpha -\hat Z)\rangle\;,\label{gen}
\end{eqnarray}
where the overbar denotes the complex conjugate,
$\d^2\lambda=\d\hbox{Re}\lambda\,\d\hbox{Im}\lambda $,
$\langle\ldots\rangle$ represents the ensemble quantum average on both signal and
image-band modes, and $\delta^{(2)}(\alpha )$ is the 
Dirac delta-function in the complex plane.  The partial trace over the
image-band mode in Eq. (\ref{gen}) can be evaluated as follows
\begin{eqnarray}
\langle e^{\lambda \hat Z^{\dag}-\bar\lambda\hat Z} \rangle
&=&\hbox{Tr}_a\left[\hat\varrho\hat D_a(\lambda) \right]
{}_b\langle 0|\hat D_b(-\bar\lambda)|0\rangle_b=
\hbox{Tr}_a\left[\hat\varrho\hat D_a(\lambda) \right]\,e^{-{1\over
2} |\lambda |^2}\nonumber\\ &\doteq& \hbox{Tr}_a\left[\hat\varrho
{\bf :}\hat D_a(\lambda) {\bf :} _{A} \right] \;,\label{part}
\end{eqnarray}
where $\hat D(\alpha)=\exp(\alpha a^{\dag}-\bar\alpha a)$ denotes the
displacement operator ($\hat D_a$ for mode $a$ and  $\hat D_b$ for
mode $b$), $|0\>_b$ represents the vacuum for mode $b$ only,
$\hat\varrho$ is the density matrix for the signal mode, and
$\hbox{\bf :}\ \hbox{\bf :}_{A}$ denotes anti-normal ordering. The
probability density {\em vs} the outcome $\alpha $ is given by
\begin{equation}  
\d^2 \alpha \,p(\alpha ,\bar \alpha )=\hbox{Tr}\left[\hat\varrho
\,\d\hat\mu(\alpha ,\bar \alpha )\right]\;,
\end{equation}
where the probability operator-valued measure (POVM) $d\hat\mu (\alpha
,\bar \alpha )$ can be written as follows
\begin{eqnarray}
\d\hat\mu (\alpha ,\bar \alpha )&=&\d^2 \alpha \,\int
{\d^2\lambda\over\pi^2}\, e^{\bar\lambda \alpha -\lambda\bar \alpha }\,
\hbox{\bf :}\hat D(\lambda)\hbox{\bf :}_{A} \nonumber\\ &=&\d^2
\alpha \,\int {\d^2\beta\over\pi}\,\int {\d^2\lambda\over\pi^2}\,
e^{\bar\lambda (\alpha -\beta)-\lambda(\bar \alpha
-\bar\beta)}\,|\beta\rangle \langle\beta|\nonumber\\ &=&{\d^2 \alpha
\over \pi}\,|\alpha \rangle\langle \alpha | \doteq\d^2 \alpha
\,\hbox{\bf :}\delta^{(2)}(\alpha -a)\hbox{\bf :}_{A} \;,
\end{eqnarray}
using the resolution of the identity in terms of coherent states $\hat
1=\int {d^2 \beta\over\pi}|\beta\rangle\langle\beta| $.  \par In a
``classical'' measurement of the function $w=f(\alpha ,\bar \alpha )$ on
the phase space, one evaluates the function $f$ of the outcome
$\alpha$ of the complex photocurrent $\hat Z$. Correspondingly, the
probability distribution of $w$ is given by the marginal probability density 
\begin{eqnarray}
p(w)=\int\d^2\alpha \,p(\alpha ,\bar \alpha )\,\delta(w-f(\alpha ,\bar
\alpha ))\;.
\end{eqnarray}
The POVM $\d\hat H_f(w)$ that provides such probability density is the
marginal POVM of $\d\hat\mu(\alpha ,\bar \alpha )$, and can be written as
follows
\begin{eqnarray}
\d\hat H_f(w)=\d w\int\d\hat\mu(\alpha ,\bar \alpha )\,\delta(w-f(\alpha
,\bar \alpha ))=\d w \,\hbox{\bf :}\delta(w-f(a,a^{\dag}))\hbox{\bf
:}_{A} \;.\label{marggen}
\end{eqnarray}
In this way one has a correspondence rule between POVM's $d\hat
H_f(w)$ and classical observables $w=f(\alpha ,\bar \alpha )$ on the
phase space $\alpha\in\mathbb C$.  
\par The quantum efficiency $\eta $ of the heterodyne detector can be
taken into account by 
introducing auxiliary vacuum field modes for both the signal and the
idler, and by rescaling the output photocurrent by an additional
factor $\eta ^{1/2}$. The overall effect resorts to a 
Gaussian convolution of the ideal POVM with variance $\Delta
^2_ {\eta}=(1-\eta )/\eta$. Then, the POVM in Eq. (\ref{marggen}) rewrites
\begin{eqnarray}
\d\hat H _f(w)=\d w \,\Gamma _{{1-\eta\over\eta}} \left[\hbox{\bf
:}\delta(w-f(a,a^{\dag}))\hbox{\bf :}_{A} \right]
\;,\label{marggeneta}
\end{eqnarray}
where $\Gamma _{\sigma ^2}$ denotes the completely positive (CP) map
that describes the effect of additional Gaussian noise of variance
$\sigma ^2$, namely
\begin{eqnarray}
\Gamma _{\sigma ^2}[\hat A]=\int {\d^2\beta\over \pi \sigma ^2}
\,e^{-{|\beta|^2\over \sigma ^2}}
\,\hat D(\beta)\hat A\hat D^{\dag }(\beta)\;,\label{dd}
\end{eqnarray}
for any operator $\hat A$.  We do not know {\it a priori} if the
measurement described by the POVM in Eq.  (\ref{marggen}) or
(\ref{marggeneta}) corresponds to an approximate quantum
measurement of some observable of the field.  We can argue that, for
example, for $f(\alpha,\bar\alpha)=|\alpha |^2$ the measurement would 
approximate the ideal detection of the number of photons
$a^{\dag}a$.  In the following we give a necessary and sufficient
condition to establish when the heterodyne POVM $\d\hat H _f(w)$
approaches the ideal quantum measurement of an observable $\hat O$ by
preamplifying the heterodyne through an ideal amplifier of $\hat O$ in
the limit of infinite amplifier gain. In the following section we
introduce the general concept of ideal amplification of an observable,
and prove that it can be always achieved by a unitary transformation.
\section{Ideal amplification of quantum observables}\label{s:ideampl}
For a given selfadjoint operator $\hat W$, the {\em ideal amplifier} of
$\hat W$ is a device that achieves the transformation
\begin{eqnarray}
{\cal A}^{(\hat W)}_g(\hat W)=g\hat W\label{amp1}\;,
\end{eqnarray}
where $g>1$ denotes the gain of the amplifier.
The transformation (\ref{amp1}) is to be regarded as the 
Heisenberg-picture evolution of the field throughout the device when
the transformation is applied to $\hat W$.  If the spectrum ${\mathbb
S}_{\hat W}$ of $\hat W$ is ${\mathbb S}_{\hat W}={\mathbb R}$ or ${\mathbb
S}_{\hat W}={\mathbb R}^+$, the evolution ${\cal A}^{(\hat W)}_g$ can be written as
follows 
\begin{eqnarray}
{\cal A}^{(\hat W)}_g{(|w\rangle\langle w|)}=g^{-1}|g^{-1}w\rangle\langle
g^{-1}w|
\label{amp2}\;,
\end{eqnarray}
where $|w\rangle$ denotes the eigenvector of $\hat W$ pertaining to
the eigenvalue $w\in{\mathbb S}_{\hat W}$. The corresponding Schr\"odinger-picture of
the evolution (\ref{amp2}) is given by the dual map
\begin{eqnarray}
{\cal A}^{\vee (\hat W)}_g{(|w\rangle\langle
w|)}=g|gw\rangle\langle g w|
\label{amp3}\;,
\end{eqnarray}
where $|w\rangle $ now has to be regarded as a (Dirac-sense)
normalized state vector. For integer spectrum ${\mathbb S}_{\hat W}={\mathbb N}$ or
${\mathbb S}_{\hat W}={\mathbb Z}$ Eq. (\ref{amp2}) rewrites as follows
\begin{eqnarray}
{\cal A}^{W}_g{(|n\rangle\langle
n|)}=|g^{-1}n\rangle\langle g^{-1}n|\,\chi_{\mathbb Z}(g^{-1}n)
\label{ampn}\;,
\end{eqnarray}
where $\chi_{\mathbb Z}(x)$ is the characteristic function on
integers, namely $\chi_{\mathbb Z}(x)=1$ for $x\in {\mathbb Z}$,
$\chi_{\mathbb Z}(x)=0$ otherwise. It is easy to check that both
Eq. (\ref{amp2}) and  
(\ref{ampn}) imply Eq. (\ref{amp1}). In the following we will 
consider only the cases of spectra ${\mathbb S}_{\hat W}={\mathbb R}$,  
${\mathbb R}^+$,  ${\mathbb N}$, ${\mathbb Z}$, as these are the only ones
that are left invariant under amplification, i. e. $g{\mathbb S}_{\hat W}
\subset{\mathbb S}_{\hat W}$ (this will exclude, for example, the case of
phase amplification \cite{hel}). Moreover, for the sake of notation, if not
explicitly written, we will assume ${\mathbb S}_{\hat W}={\mathbb R}$. 
\par Among all possible extensions of the amplification map
(\ref{amp3}) to all state vectors, the following ones are physically
meaningful 
\begin{eqnarray}
&&{\cal A}^{\vee (\hat W)}_g{(|w\rangle\langle
w'|)}=g|gw\rangle\langle gw'| 
\label{ampd1}\;,\\
&&{\cal A}^{\vee (\hat W)}_g{(|w\rangle\langle
w'|)}=g|gw\rangle\langle gw'|\delta(w-w')
\label{ampd2}\;.
\end{eqnarray}
In fact, both maps in Eqs. (\ref{ampd1}) and (\ref{ampd2}) are linear normal
{\em completely positive} (CP) maps, and hence they can be realized through
a unitary transformations on an extended Hilbert space \cite{ozw}. The
proof runs as follows. The map $A$ is completely positive normal
if and only if one has
\begin{eqnarray}
\sum_{i,j=1}^{n}\langle\xi_i|A^{\vee}(|\eta_i\rangle\langle\eta_j|)
|\xi_j\rangle \geq 0\;
\end{eqnarray} 
for all finite sequence of vectors $\{|\eta_i\>\}$ and
$\{|\xi_i\>\}$. Upon expanding $|\eta_i\>$ and $|\xi_i\>$ on the
orthonormal basis $\{|w\>\}$, for the map (\ref{ampd1}) one has 
\begin{eqnarray}
&&\sum_{i,j=1}^{n}\langle\xi_i|A_g^{\vee (\hat W)}
(|\eta_i\rangle\langle\eta_j|)|\xi_j\rangle =g\int
dw_1\,dw_2\,dw_3\,dw_4\,\langle w_1|gw_2\rangle\langle gw_3|w_4\rangle
\nonumber\\ &&\times \sum_{i,j=1}^n \bar\xi _i(w_1)
\eta_i(w_2)\bar\eta _j(w_3) \xi _j(w_4)\nonumber \\
&&=g\left|\sum_{i=1}^n \int dw\,dw'\langle w|gw'\rangle \bar\xi _i(w)
\eta_i(w')\right|^2\geq 0\;,
\end{eqnarray}
whereas for the map (\ref{ampd2}) one has
\begin{eqnarray}
&&\sum_{i,j=1}^{n}\langle\xi_i|A_g^{\vee (\hat W)}
(|\eta_i\rangle\langle\eta_j|)|\xi_j\rangle \nonumber \\= &&g\int
_{-\infty}^{+\infty}{d\lambda \over 2\pi} \left|\sum_{i=1}^n \int
dw\,dw'\langle w|gw'\rangle \bar\xi _i(w) \eta_i(w')e^{i\lambda
w'}\right|^2\geq 0\;.
\end{eqnarray}
\par In the Schr\"odinger picture the two maps (\ref{ampd1}) and
(\ref{ampd2}) are achieved by the following unitary transformations in
an extended Hilbert space 
\begin{eqnarray}
&&\hat U _g|w\rangle\otimes|\psi\rangle =g^{\frac
12}|gw\rangle\otimes|\psi '\rangle\label{cohamp}\\ &&\hat U
_g|w\rangle\otimes|\psi\rangle =g^{\frac 12}|gw\rangle\otimes|\psi'
(w)\rangle\;.
\label{ncohamp}
\end{eqnarray}
with $\langle \psi'(w_1)|\psi '(w_2)\rangle =\delta(w_1-w_2)$.  
Eqs. (\ref{ampd1}) and (\ref{ampd2}) are obtained by
Eqs. (\ref{cohamp}) and (\ref{ncohamp}) when the evolution 
is viewed as restricted to the signal mode only, namely
\begin{eqnarray}
{\cal A}^{\vee (\hat W)}_g(\hat\varrho)=\langle\psi|
\hat U_g \hat\varrho \otimes 1 \hat U_g^{\dag}|\psi\rangle
\label{dual}\;.
\end{eqnarray}
We name the device corresponding to Eq. (\ref{cohamp}) an {\em ideal
coherence-preserving quantum amplifier} of $\hat W$, because it
achieves the ideal amplification of $\hat W$ without measuring $\hat
W$ ($\psi'$ does not depend on $w$; for $\psi'=\psi$ the device is
``passive''). On the other hand, the transformation (\ref{ncohamp})
achieves the ideal amplification of $\hat W$ by measuring $\hat W$,
then performing the processing $w\to gw$, and finally preparing the
state $|gw\>$. The measurement stage is the one which is responsible
for the vanishing of all off-diagonal elements in Eq. (\ref{ampd2}).  
(Eq. (\ref{dual}) together with Eq. (\ref{cohamp}) and (\ref{ncohamp})
imply Eqs. (\ref{ampd1}) and (\ref{ampd2}) also for a nonorthogonal
set $\{|w\rangle\}$, however, generally not when
$\<\psi'_A(w_2)|\psi'_A(w_1)\rangle\neq 0$ for $w_1\neq w_2$). Since we
want to exploit the ideal amplification of $\hat W$ in order to
achieve its ideal quantum measurement, we will consider only the
coherence-preserving quantum amplification in Eq. (\ref{ampd1}) or
(\ref{cohamp}), since the other kind of amplifier needs by itself the
ideal measurement of $\hat W$. 
\section{Approaching ideal quantum measurements by 
preamplified heterodyning}\label{s:criter}
Let $\d\hat H_f(u)$ be the POVM pertaining
to the heterodyne measurement of the function $f(\alpha ,\bar\alpha)$
of the field, and let consider a preamplified heterodyne detection
scheme corresponding to the following procedure:
\begin{enumerate}
\item the signal mode of the field is amplified by an ideal
amplifier for $\hat W$ with gain $g$; 
\item the field is heterodyne
detected and the function $f$ is evaluated;
\item the final result is rescaled by a factor $g$.
\end{enumerate}
The above procedure corresponds to the following transformation
\begin{eqnarray}
\d\hat H_f(u)\longrightarrow 
{\cal A}_g^{(\hat W)}[\d\hat H_f(gu)]\;.
\label{presc}
\end{eqnarray}
We say that the preamplified heterodyne detection of the function
$f$ of the field approaches the ideal quantum measurement of the
observable $\hat W$ in the limit of infinite gain $g$ if
\begin{eqnarray}
\lim_{g\to\infty}{\cal A}_g^{(\hat W)}[\d\hat H_f(gu)]=\d u\,\delta(u-\hat W)\;,
\label{limit}
\end{eqnarray}
where the limit is to be regarded in the weak sense (i. e. for matrix
elements) and the operator Dirac delta explicitly writes as follows 
\begin{eqnarray}
\delta(u-\hat W)=\int_{{\mathbb S}_{\hat W}}\d w|w\>\< w|\delta(u-w)\;,
\end{eqnarray}
and the integral is to be understood as a sum for discrete spectrum
${\mathbb S}_{\hat W}$. A necessary and sufficient condition for validity of
Eq. (\ref{limit}) is the following 
\begin{eqnarray}
\lim_{g\to\infty}\int{\cal A}_g^{(\hat W)}[\d\hat H_f(gu)]u^l=\hat
W^l\;,\qquad l=0,1,2,\ldots\;,\label{nec-suff}
\end{eqnarray}
where again the limit holds for expectations on any state. One can prove
that condition (\ref{nec-suff}) is necessary---i.e. Eq. (\ref{limit})
implies Eq. (\ref{nec-suff})---by simply substituting Eq. (\ref{limit}) into
Eq. (\ref{nec-suff}), and exchanging the integral with the limit. On
the other hand Eq. (\ref{nec-suff}) implies
\begin{eqnarray}
\lim_{g\to\infty}\int{\cal A}_g^{(\hat W)}[\d\hat H_f(gu)]\exp(iku)=
\exp(ik\hat W)\;,
\end{eqnarray}
and taking the Fourier transform of both sides of the last identity
one finds Eq. (\ref{limit}), proving that Eq. (\ref{nec-suff}) is also
a sufficient condition. Another sufficient condition in a form more
convenient than Eq. (\ref{nec-suff}) is the following 
\begin{eqnarray}
\int\d\hat H_f(u)u^l=\hat W^l+o(\hat W^l)\;,
\label{suff2}
\end{eqnarray}
where $o(g(x))$ is an asymptotic notation equivalent to the vanishing
of the limit $\lim_{x\to\infty} o(g(x))/g(x)=0$ \cite{estrada},
whereas, for an operator $\hat V$, by $o(\hat V)$ we mean
$\lim_{\kappa\to\infty}\kappa^{-1}o(\kappa\hat V)=0$ in the
weak sense. In fact, by amplifying both sides of Eq. (\ref{suff2}) and
rescaling the variable $u$ by the gain $g$ one obtains 
\begin{eqnarray}
\int{\cal A}_g^{(\hat W)}[\d\hat H_f(gu)]u^l=\hat
W^l+g^{-l}o(g^l\hat W^l)\;,
\end{eqnarray}
which implies Eq. (\ref{nec-suff}).    
\section{Two examples}\label{s:examples}
In this section we show that condition (\ref{nec-suff}) holds for both
the photon number $\hat W=a^{\dag}a$ and the quadrature $W=\hbox{Re}
(a\,e^ {-i\phi})$, corresponding to the functions of the field
$f(\alpha,\bar\alpha)=|\alpha|^2$ and $f(\alpha,\bar\alpha)=\hbox{Re}
(\alpha\,e^{-i\phi})$ respectively. This means that both the
quadrature and the photon number operators can be ideally measured
through the preamplified heterodyne detection scheme in the limit of
infinite gain. We also show that in both cases the detection scheme is
robust to nonunit quantum efficiency of the heterodyne detector.
\subsection{Measurement of the quadrature}
The POVM $\d\hat H (x)$ that corresponds to the function
$f(\alpha,\bar\alpha)\!=\!\hbox{Re} (\alpha\,e^{-i\phi})$ of the field is
given by
\begin{eqnarray}
&&\d\hat H_f (x)=\d x \,\hbox{\bf :} \delta \left(x-{1\over2}(a^{\dag}\,e^{i\phi
}+ a\,e^{-i\phi })\right)\hbox{\bf :}_{A}= dx\,\int {du\over
{2\pi}}e^{iu(x-\hat X _\phi)}\,e^{-\frac 18 u^2} \nonumber\\&&
=dx\,\sqrt{2\over \pi}\,e^{-2(\hat X_\phi -x)^2}\label{pq}\;.
\end{eqnarray}
Nonunit quantum efficiency introduces additive Gaussian
noise and replaces the POVM (\ref{pq}) with the following one
\begin{eqnarray}
\d\hat H_{f}(x)=\d x\,\sqrt{2\eta\over \pi
(2-\eta)}\,e^{-{2\eta\over 2-\eta} (\hat X_\phi -x)^2}\label{pqeta}\;.
\end{eqnarray}
We can see that the POVM in Eq. (\ref{pqeta}) satisfies the
sufficient condition (\ref{suff2}) for approaching the ideal
quantum measurement of $\hat X_{\phi}$. In fact, the moments of the
POVM (\ref{pqeta}) are given by  
\begin{equation}
\int\d\hat H_f(x)x^l=\int_{-\infty}^{+\infty}\d x\,
\sqrt{2\eta\over \pi (2-\eta)}\,e^{-{2\eta\over 2-\eta}x^2}
(\hat X_\phi +x)^l=\hat X_{\phi}^l +O(\hat X_{\phi}^{l-2})\;,
\end{equation}
where $O(g(x))$ is the customary asymptotic notation equivalent to the
condition $\lim_{x\to\infty} O(g(x))/g(x)<\infty$ \cite{estrada}, 
implying that $O(\hat X^{l-2})\equiv o(\hat X^l)$. On the other
hand, one can directly verify the limit in Eq. (\ref{limit}) as follows
\begin{eqnarray}
{\cal A}_g^{\hat X_{\phi}}[\d\hat H_f(gx)]=\d x\,\sqrt{2 g^2\eta\over \pi
(2-\eta)}\,e^{-{2 g^2\eta\over 2-\eta} (\hat X_\phi -x)^2}
\stackrel {g\rightarrow \infty}{\longrightarrow}
\d x\,\delta(\hat X_\phi- x)\;.
\end{eqnarray}
\par The ideal amplification of the quadrature operator $\hat X_\phi $
is achieved by means of a phase-sensitive amplifier \cite{yuen5a,yuen6a}
which rescales the couple of conjugated quadratures as follows
\begin{eqnarray}
\hat X_\phi \rightarrow \frac 1g\hat X_\phi\,,\qquad
\hat X_{\phi+\frac \pi 2} \rightarrow
g\hat X_{\phi+\frac \pi 2}\label{xy}\;,
\end{eqnarray}
$g$ being the gain at the amplifier. The Heisenberg transformations in
Eq. (\ref{xy}) are achieved by the unitary operator
\begin{eqnarray}
\hat U_g=\exp[-i\log g (\hat X_{\phi }\hat X_{\phi+\frac \pi 2}-
\hat X_{\phi+\frac \pi 2}\hat X_{\phi })]\;.
\end{eqnarray}
\subsection{Measurement of the photon number}
The case of the ideal measurement of the photon number $a^{\dag}a$
through preamplified heterodyning is more interesting than the case of
the quadrature $\hat X_\phi$, because here the amplification not only
removes the excess noise due to the quantum measurement, but also
changes the spectrum, from continuous to discrete. We consider
the POVM that corresponds to heterodyning the function $f(\alpha
,\bar\alpha)=|\alpha |^2$ of the field. This can be written as follows
\begin{eqnarray}
&&\d\hat H_f (h)=\d h\,\hbox{\bf :}\delta(h-a^{\dag}a))\hbox{\bf :}_{A}=
\d h\,\int {\d u\over {2\pi}}\,e^{-iuh}\,\sum_{n=0}^{\infty}
(iu)^n\,a^na^{\dag n} \nonumber\\&&=\d h\,\int {\d u\over
{2\pi}}\,e^{-iuh}\, \sum_{n=0}^{\infty} (iu)^n\,{a^{\dag }a+n\choose
n} =\d h\,\int {\d u\over {2\pi}}\,e^{-iuh}\, (1-iu)^{-a^{\dag}a-1}
\nonumber\\&&=\d h\,e^{-h}\,{h^{a^{\dag}a}\over{(a^{\dag}a)!}}\label{pomn}\;.
\end{eqnarray}
The POVM in Eq. (\ref{pomn}) satisfies the sufficient condition
(\ref{suff2}). In fact, one has 
\begin{eqnarray}
&&\int\d h\, e^{-h}\,\frac{h^{a^{\dag}a+l}}{(a^{\dag}a)!}=\frac{(a^{\dag
}a+l)!}{(a^{\dag }a)!}\nonumber\\ &&=(-)^l\sum_{k=0}^ls_{l+1}^{(k+1)}(-a^{\dag
}a)^k=(a^{\dag }a)^l+O[(a^{\dag }a)^{l-1}]\;,\label{stirl}
\end{eqnarray}
where $s_{l}^{(k)}$ denotes a Stirling number of the first kind.
Hence, if the field is amplified through an ideal photon number
amplifier \cite{pna1,yuen6a,pna11} and then heterodyne detected, 
in the limit of infinite gain the scheme achieves ideal photon number
detection. Indeed, using the ideal photon number amplification map
\cite{pna2,pna3} 
\begin{eqnarray}
a^{\dag }a \longrightarrow \hat V^{\dag }\,a^{\dag }a\,\hat
V=g\,a^{\dag }a\;,
\end{eqnarray}
with the isometry $\hat V$ given by
\begin{eqnarray}
\hat V=\sum_{n=0}^{\infty }|gn\rangle \langle n|\;,
\end{eqnarray}
one obtains the preamplified POVM
\begin{eqnarray}
{\cal A}_g^{a^{\dag}a}[\d\hat H_f(gh)]=\hat V^{\dag }\,\d\hat H_f(gh) \,\hat
V=\d h\,g\,e^{-gh}\sum_{n=0}^{\infty }\frac{(gh)^{gn}}{(gn)!}
|n\rangle\langle n|\;.\label{pren}
\end{eqnarray}
In the limit of infinite gain $g\rightarrow \infty $ the POVM in
Eq. (\ref{pren}) achieves the ideal POVM for the photon-number operator
measurement. This can be shown as follows. Upon writing the POVM
(\ref{pren}) in the form
\begin{eqnarray}
{\cal A}_g^{a^{\dag}a}[\d\hat H_f(gh)]=\d h\sum_{n=0}^{\infty }
p_n^{(g)}(h)|n\rangle\langle n|\;,
\end{eqnarray}
we need to show that the function
\begin{eqnarray} 
p_n^{(g)}(h)=g\,e^{-gh}\frac{(gh)^{gn}}{(gn)!}
\;,\label{png}
\end{eqnarray}
approaches a Dirac delta-comb over integer values $h\in{\mathbb N}$.
Using the Stirling's inequality
\begin{eqnarray} 
\sqrt{2\pi n}\left(\frac n e\right )^n 
<n!<\sqrt{2\pi n}\left(\frac n e\right)^n \left(1+\frac{1}{12n-1}\right)
\;,\label{stir}
\end{eqnarray}
one obtains
\begin{eqnarray} 
\gamma _n^{(g)}(h)
\left(1+\frac{1}{12gn-1} \right)^{-1}<p_n^{(g)}(h)<
\gamma _n^{(g)}(h) 
\;,\label{carab}
\end{eqnarray}
where
\begin{eqnarray} 
\gamma _n^{(g)}(h)=\frac{1}{\sqrt{2\pi g^{-1}n}}\exp \left[gn\left(
1-\frac hn+\log\frac hn\right)\right]
\;.\label{gam}
\end{eqnarray}
From the inequality $\log x\leq x-1$ (with equality
iff $x=1$) it follows that
\begin{eqnarray} 
\lim _{g\rightarrow\infty} \gamma_n^{(g)}(h)=
\left\{\begin{array}{cc}
0 & h\neq n \\
+\infty & h=n \end{array} 
\right. \;,
\end{eqnarray}
and hence, from Eq. (\ref{carab}), one has
\begin{eqnarray} 
\lim _{g\rightarrow\infty} p_n^{(g)}(h)= \left\{\begin{array}{cc} 0 &
h\neq n \\ +\infty & h=n \end{array} \right. \;.
\end{eqnarray}
Moreover, from the expansion for $h$ near to $n$
\begin{eqnarray}
1-\frac hn +\log \frac hn=-\frac 12 \left(1-\frac hn\right)^2 +
O\left(\left (1-\frac hn\right)^3\right)\;,
\end{eqnarray}
one has the Gaussian asymptotic approximation for $g\rightarrow\infty$
\begin{eqnarray}
p_n^{(g)}(h)\simeq\frac{1}{\sqrt{2\pi g^{-1}n}} \exp
\left[-\frac{\left( h-n\right)^2}{2g^{-1}n}\right]
\stackrel {g\rightarrow \infty} {\longrightarrow} 
\delta (h-n)\;,\label{asym}
\end{eqnarray}
which proves the statement.
\par
\begin{figure}[hbt]
\vskip 18pt
\begin{center}
\epsfxsize=.6\hsize\leavevmode\epsffile{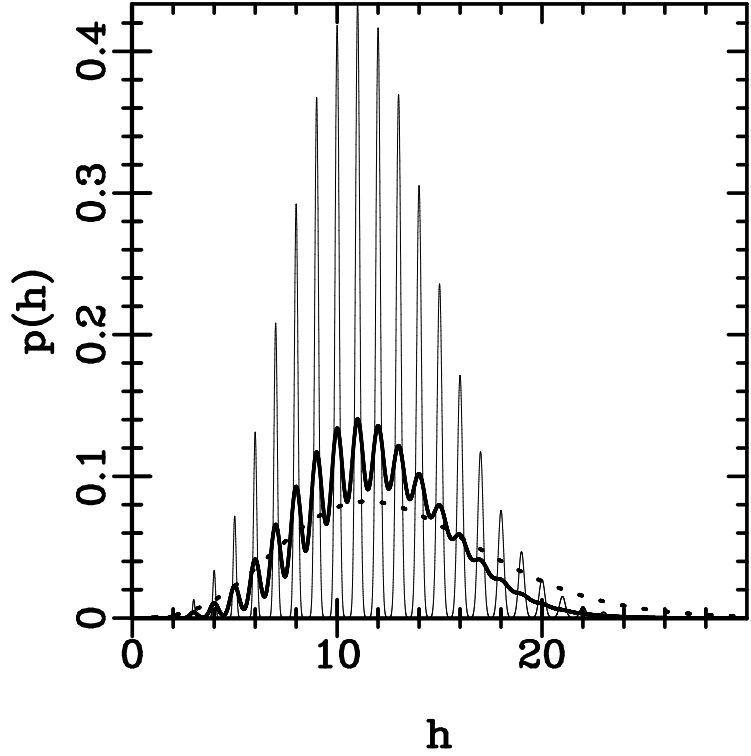}
\end{center}
\vfill
\caption{Probability density $p(h)$ for a coherent state with mean
photon number $\langle a^{\dag}a\rangle =12$ obtained through
heterodyne detection of $f(\alpha ,\bar\alpha)=|\alpha |^2$, preamplified 
by an ideal photon number amplifier. Different line-style denote different
value of the gain $g$ at the amplifier: the dashed line corresponds to
$g=1$ (no amplification); the thick line corresponds to $g=10^2$; the
thin line to $g=10^3$.} 
\label{f:pn}
\end{figure}

 In Fig. \ref{f:pn} we show the probability distribution
of the outcome $h=|\alpha |^2$ form preamplified heterodyne detection
of a coherent state, for different values of the amplifier gain $g$.
Notice the emergence of a discrete spectrum from a continuous one for
increasingly large gains, in agreement with Eq. (\ref{asym}).   
\par It is easy to show that the preamplified heterodyne detection
scheme is robust to nonunit quantum efficiency also in the present
case of measuring $a^{\dag}a$. In fact, the sufficient condition  
(\ref{suff2}) is still satisfied for nonunit quantum efficiency, as
one can check through Eqs. (\ref{dd}) and (\ref{stirl}) as follows 
\begin{eqnarray}
&&\int {\d^2\beta\over \pi }\frac{\eta
}{1-\eta}\,e^{-\frac{\eta}{1-\eta} {|\beta|^2}} \, \hat
D(\beta)\left\{(a^\dag a)^l+O[(a^\dag a)^{l-1}]\right\}\hat
D^{\dag }(\beta)\nonumber\\ &&=\int {\d^2\beta\over \pi }\frac{\eta
}{1-\eta}\,e^{-\frac{\eta}{1-\eta} {|\beta|^2}}
\,\left\{[(a^\dag-\overline{\beta})(a-\beta)]^l
+O[((a^\dag-\overline{\beta})(a-\beta))^{l-1}]\right\}
\nonumber\\ &&= (a^\dag a)^l+O[(a^\dag a)^{l-1}]\;.\label{due}
\end{eqnarray}
\section{A counterexample}\label{s:counterex}
The necessary and sufficient condition (\ref{nec-suff}) establishes
when a self-adjoint operator $\hat W$ is approximated by the classical
observable $f$ using a preamplified heterodyne scheme. One could now
address the inverse problem, namely: Given a self-adjoint operator
$\hat W$ is it possible to find a function of the field such that the
preamplified heterodyne measurement approximates the measurement of
$\hat W\,$? As we have shown in the previous section, this is
certainly true for $\hat X_\phi $ and $a^\dag a$. For a generic
observable $\hat W$, the problem becomes very difficult.  However, on
the basis of a counterexample, we will prove that the inverse problem
has no solution for some operator $\hat W\,$, namely there are
observables which cannot be measured through the
preamplified heterodyne detection scheme.   
\par Consider the operator
\begin{eqnarray} 
\hat K\equiv -\frac i2(a^2-a^{\dag 2})=\hat X\hat Y+\hat Y\hat X
\;,
\end{eqnarray}
where $\hat X$ and $\hat Y$ are the conjugated quadratures $\hat
X\equiv\hat X_0$ and $\hat Y=\hat X_{\pi/2}$. We show that there is no
polynomial function of the field that satisfies either the necessary
condition (\ref{nec-suff}).
\par In order to construct the 
CP amplification map for $\hat K$, one has to find the eigenstates of 
$\hat K$. These are given in Ref. \cite{boll}, and here we report
them. One has
\begin{eqnarray} 
\hat K|\psi ^\mu _{\pm}\rangle =\mu |\psi ^\mu _{\pm}\rangle \;,
\end{eqnarray}  
with
\begin{eqnarray} 
\psi ^\mu _\pm (x)\doteq \langle x|\psi ^\mu _{\pm}\rangle
=\frac{1}{\sqrt{2\pi}}\,|x|^{i\mu-\frac 12}\theta(\pm x)\;,
\end{eqnarray}  
where $|x\rangle $ denotes the eigenvector of the quadrature
$\hat X$, and $\theta(x)$ is the customary step-function ($\theta(x)=1$
for $x>0$, $\theta(x)=1/2$ for $x=0$, $\theta(x)=0$ for for $x<0$).
The vectors $|\psi _s^\mu \rangle$ form a complete orthonormal set
\begin{eqnarray} 
\langle\psi _r^\mu |\psi _s^\nu \rangle =\delta _{rs}\delta(\mu-\nu)
\;.
\end{eqnarray}  
The amplification of $\hat K$ is achieved by the unitary operator 
$\hat U_g$ satisfying the relations
\begin{eqnarray}
\hat U ^{\dag }_g \hat K \hat U_g=g\hat K\;,\qquad
\hat U _g |\psi _s^\mu \rangle=g^{\frac 12}|\psi _s^{g\mu } 
\rangle\,. \label{uk1}
\end{eqnarray}
In terms of the eigenvectors of $\hat K$ the unitary operator $\hat
U_g$ has the form
\begin{eqnarray}
\hat U _g=\sum _{s=\pm}\int_{-\infty}^{+\infty}\d\mu\,g^{\frac 12}
|\psi _s^{g\mu } 
\rangle \langle \psi _s^{\mu } |
=g^{\frac 12}\int _{-\infty}^{+\infty}\d x\, |x|^{\frac 12(g-1)}
|x\rangle\langle 
x^{*g}|\label{uk2}
\;,
\end{eqnarray}  
where in the last identity in Eq. (\ref{uk2}) we have written
$\hat U_g$ in terms of the eigenstates $|x\rangle $ of the quadrature
$\hat X$, upon introducing  the notation 
\begin{eqnarray}
x^{*g}\equiv x|x|^{g-1}=\hbox{sgn}(x)|x|^g \;,
\end{eqnarray}
where $\hbox{sgn}(x)$ denotes the customary sign function.
The analytic form (\ref{uk2}) of $\hat U_g$ is derived as follows
\begin{eqnarray}
&&\hat U _g = g^{\frac 12}\sum _{s=\pm}\int_{-\infty}^{+\infty}\d\mu\,
|\psi _s^{g\mu } \rangle \langle \psi _s^{\mu } |\nonumber \\&&
=g^{\frac 12}\int_{-\infty}^{+\infty} \d x\int_{-\infty}^{+\infty} \d x'
|x'\rangle\langle x| \sum _{s=\pm}\int_{-\infty}^{+\infty}d\mu\,
\psi_s^{g\mu }(x') \bar\psi_s^{\mu }(x)\nonumber \\&& =g^{\frac
12}\int_{-\infty}^{+\infty} \d x\int_{-\infty}^{+\infty} \d x'
|x'\rangle\langle x| \,|x'|^{\frac 12(g-1)} \sum
_{s=\pm}\int_{-\infty}^{+\infty}d\mu\, \psi_s^{\mu }(x'^{*g})
\bar\psi_s^{\mu }(x)\nonumber \\&& =g^{\frac 12}\int
_{-\infty}^{+\infty}\d x\, |x|^{\frac 12(g-1)}|x\rangle\langle x^{*g}|
\;.\label{uk3}
\end{eqnarray}  
The Heisenberg evolution of the conjugated quadratures $\hat X $ and
$\hat Y$ by the amplification $\hat U_g$ can be evaluated through the
following steps
\begin{eqnarray}
\hat U^{\dag }_g\hat X\hat U_g &=& g\int _{-\infty}^{+\infty}\d x\, 
\int _{-\infty}^{+\infty}\d x'\, 
x|xx'|^{\frac 12(g-1)}|x^{*g}\rangle\langle x|x'\rangle\langle
x'^{*g}|\nonumber\\&=& g\int _{-\infty}^{+\infty}\d x\,
x^{*g}|x^{*g}\rangle\langle x^{*g}|=\hat X^{*\frac 1g}
\label{uxu} \;;
\\
\hat U^{\dag }_g\hat Y\hat U_g&=&\hat U^{\dag }_g \int
_{-\infty}^{+\infty}\d x\, |x\rangle
\left(-\frac i2 \partial _x\right)\langle x| \hat U_g\nonumber\\&=&
g\int _{-\infty}^{+\infty}\d x\, |x|^{\frac 12(g-1)} |x^{*g}\rangle
\left(-\frac i2 \partial _x\right) \langle x^{*g}| |x|^{\frac
12(g-1)}\nonumber\\
&=& -\frac i4 (g-1)\hat X^{*(-\frac 1g)}+
\int _{-\infty}^{+\infty}du\, |u\rangle \left(-\frac i2 |u|^{1-\frac
1g}\partial _u\right) \langle u|\nonumber\\&=&-\frac i4 (g-1)\hat
X^{*(-\frac 1g)}+g\hat X^{*(-\frac 1g)}\hat X\hat Y
\nonumber \\
&=&\hat X^{*(-\frac 1g)}\left(\frac 12
g\hat K +\frac i4\right)= \left(\frac 12 g\hat K -\frac i4\right)\hat
X^{*(-\frac 1g)}
\label{uyu} \;.
\end{eqnarray}  
For what follows we also need to evaluate the Heisenberg evolution
of the operator $\hat X^2 +\hat Y^2 =a^{\dag }a +\frac 12$.  From
Eqs. (\ref{uxu}-\ref{uyu}) one has
\begin{eqnarray}
\hat U^{\dag }_g\left(a^{\dag } a+\frac 12\right) \hat U_g&=&|\hat
X|^{\frac 2g}+\frac 14 \left(g\hat K-\frac i2\right)|\hat X|^{(-\frac
2g)} \left(g\hat K+\frac i2\right) \nonumber\\&=& |\hat X|^{\frac
2g}+\frac 14 \hat X^{*(-\frac 1g)} \left(g^2\hat K^2+\frac
14\right)\hat X^{*(-\frac 1g)}
\label{u2u} \;.
\end{eqnarray}  
Now, let us consider a quadratic function of the field $f(\alpha ,
\bar\alpha)=-i(\alpha ^2 -\bar\alpha ^2 +ic|\alpha |^2)/2$, $c$ 
an arbitrary constant, and let us evaluate the corresponding POVM
$\d\hat H_f(u)$ pertaining to heterodyne detection of the function $f$
of the field. From Eq. (\ref{marggen}) one has
\begin{eqnarray}
&&\d\hat H_f(u)=\d u\,\hbox{\bf :}
\,\delta(u-f(a,a^{\dag}))\hbox{\bf :}_{A}\nonumber\\
&&=\d u\,\int_{-\infty}^{+\infty}\frac
{\d\lambda}{2\pi}\,e^{-i\lambda u}\, e^{\lambda
\frac{a^2}{2}}\,\hbox{\bf :}e^{i\lambda \frac c2 a^{\dag }a} \hbox{\bf
:}_A\,e^{-\lambda \frac{a^{\dag 2}}{2}}\nonumber\\ &&=
\d u\,\int_{-\infty}^{+\infty}\frac {\d\lambda}{2\pi}\,e^{-i\lambda u}\,
e^{\lambda \frac{a^2}{2}}\,\left(1-i\lambda \frac c2\right)^{-(a^{\dag
}a+ \frac 12)}\, e^{-\lambda \frac{a^{\dag 2}}{2}}\, \left(1-i\lambda
\frac c2\right)^{-\frac 12 }\!\!\!\!\! ,
\label{k1}
\end{eqnarray}
where we used the relation
\begin{eqnarray}
\hbox{\bf :}\,e^{za^{\dag }a}\hbox{\bf :}_A =\sum_{n=0}^{\infty}
\frac {z^n}{n!}\, a^n a^{\dag n}=\sum_{n=0}^{\infty} z^n {a^{\dag }a+n
\choose n}=(1-z)^{-a^{\dag }a-1}
\label{ord} \;.
\end{eqnarray}  
The product of operators in the last equality of Eq. (\ref{k1}) can be
recast in the form of a single exponential function using the
Baker-Campbell-Hausdorff (BCH) formula for the {\em su}(1,1) algebra
[see Appendix]. According to the prescription in Eq. (\ref{presc}), we
need to evaluate the preamplified POVM
\begin{eqnarray}
{\cal A}_g^{(\hat K)}[\d\hat H_f(gu)]\equiv g\,\hat U^{\dag}_g \,(\d\hat H_f(gu))\,\hat U_g 
\label{pre}\;,
\end{eqnarray}
in the limit of infinite gain $g\rightarrow\infty $.  As shown in the
Appendix, for the leading term in $g$ one has 
\begin{eqnarray}
{\cal A}_g^{(\hat K)}[\d\hat H_f(gu)]
&=&\d u\int \frac{d\lambda}{2\pi}\, \exp\left( -i\lambda
u+i\lambda\hat K +\frac 18 i\lambda cg\hat K^2\right)\nonumber\\
&\times&\exp\left[-\frac 18\lambda ^2\left(1+ \frac {c^2}{4}\right)
\hat K^2\right]\;,\qquad g\gg 1  
\label{pre2}\;.
\end{eqnarray}
The preamplified POVM in the limit of infinite gain writes as follows
\begin{eqnarray}
&&{\cal A}_g^{(\hat K)}[\d\hat H_f(gu)]\nonumber\\
&&\stackrel {g\rightarrow\infty}{\longrightarrow }
\d u\int \frac{d\lambda}{2\pi} \,\exp\left( -i\lambda
u+i\lambda\hat K +{i\over8}\lambda g c\hat K^2\right) \exp\left(-\frac
18\lambda ^2\hat K^2\right) \nonumber\\ &&=\d u\sqrt{\frac {2}{\pi\hat
K^2}} \exp\left(-\frac{2(\hat K +{1\over8}g c \hat K^2 -u)^2}{\hat K^2}
\right)
\label{pre3}\;.
\end{eqnarray}  
The POVM in Eq. (\ref{pre3}) satisfies the necessary condition
(\ref{nec-suff}) for $l=0,1$ upon choosing $c=0$. However, the same
condition for $l=2$ is not satisfied, because one has
\begin{eqnarray}
\int du\,u^2\,\sqrt{\frac {2}{\pi\hat K^2}} \exp\left(-\frac{2(\hat K
-u)^2}{\hat K^2}\right)= \frac 54 \hat K^2
\label{no2}\;.
\end{eqnarray}  
Therefore, there is no quadratic function $f(\alpha,\bar\alpha)$ of
the field that allows to approximate the ideal quantum measurement of
the operator $\hat K=-i (a^2 -a^{\dag 2})/2$. It is clear that also
higher-degree polynomial functions of the field cannot satisfy
condition (\ref{nec-suff}), since in such case higher powers in $a^{\dag}$ and
$a$ will appear in Eq. (\ref{k1}) and the BCH formula will have no
longer closed form. 
In conclusion of this section we notice that Eq. (\ref{pre3}) for
$c=0$ can also be easily obtained by the following formal asymptotic
analysis 
\begin{eqnarray}
&&\d\hat H_f(gu)=
\d u\,\int_{-\infty}^{+\infty}\frac
{\d\lambda}{2\pi}\,e^{-i\lambda u}\, 
e^{g^{-1}\lambda\frac{a^2}{2}}e^{g^{-1}-\lambda\frac{a^{\dag 2}}{2}}
\nonumber\\ &&=\d u\,\int_{-\infty}^{+\infty}\,e^{-i\lambda u}\,\exp\left\{
ig^{-1}\lambda\frac 12(a^2-a^{\dag 2})-
\frac 18g^{-2}\lambda^2[a^2,a^{\dag 2}]+O(g^{-3})\right\}\nonumber\\
&&=\d u\,\int_{-\infty}^{+\infty}\,e^{-i\lambda u}\,\exp\left\{
ig^{-1}\lambda\hat K-\frac 12g^{-2}\lambda^2\left(a^{\dag}a+\frac
12\right)+O(g^{-3})\right\}\;.
\label{last}
\end{eqnarray}
By amplifying the first and last members of Eq. (\ref{last}) and using
Eq. (\ref{u2u}) one has
\begin{eqnarray}
&&{\cal A}_g^{(\hat K)}[\d\hat H_f(gu)]\nonumber\\
&&=\d u\,\int_{-\infty}^{+\infty}\,e^{-i\lambda u}\,\exp\left\{
i\lambda\hat K-\frac 18\lambda^2\hat X^{*(-\frac 1g)}\left(\hat K^2+\frac
1{4g^2}\right)\hat X^{*(-\frac 1g)}+O(g^{-3})\right\}
\nonumber\\&& 
=\d u\,\int_{-\infty}^{+\infty}\,e^{-i\lambda u}\,\exp\left\{
i\lambda\hat K-\frac 18\lambda^2\hat K^2+O(g^{-1})\right\}\;,
\end{eqnarray}
namely Eq. (\ref{pre3}). 
\section{Conclusions}\label{s:conclu}
One may think that the heterodyne detector could be regarded as a
universal detector, as it achieves the ideal measurement of the field
operator $a$, and hence, in principle, it should achieve the
measurement of any operator $\hat O=\hat O(a,a^{\dag})$ of the
field. However, due to the fact that the measurement of $a$
corresponds to a joint measurement of two 
noncommuting conjugate observables, an intrinsic unavoidable 3dB noise
is added to the measurement, even in the ideal case. We have
considered the possibility of reducing such noise by means of a
suitable ideal preamplification of $\hat O$, which we have shown to be
feasible through a unitary transformation. We have shown that in the
limit of infinite gain such preamplified heterodyne detection scheme
can achieve the ideal measurement of $a^{\dag}a$ and $\hat X_{\phi}$,
even for nonunit quantum efficiency, also realizing the transition
from continuous to discrete spectrum in the case of the 
operator $a^{\dag}a$. However, the scheme does not work for arbitrary
operator, and, as a counterexample, we proved that the ideal
measurement cannot be achieved even for the simple quadratic 
form $\hat K=i(a^{\dag 2}-a^2)/2$, apparently with no
simple physical explanation other than the algebraic nature of the
operator $\hat K$ itself and its ideal amplification
map. In the present study we have
seen some of the problems that would appear in building 
a universal detection machine, and we hope that this work will shed new
light on the route for achieving such a challenging task.
\section*{Acknowledgements}
This work is supported by the Italian Ministero dell'Universit\`a e
della Ricerca Scientifica e Tecnologica under the program {\em 
Amplificazione e rivelazione di radiazione quantistica}, and in part
by the Office of Naval Research. 
\section*{Appendix on the BCH formula}
Upon defining $k_+=\frac 12 a^{\dag 2}$, $k_-=\frac 12 a^{2}$, and
$k_3=\frac 12( a^{\dag }a +\frac 12)$, one recognizes the following
commutation rules for the {\em su}(1,1) algebra
\begin{eqnarray}
[\hat k_+,\hat k_-]=-2\hat k_3\;,\qquad [\hat k_3,\hat k_\pm]=\pm \hat
k_\pm
\label{comm} \;.
\end{eqnarray}   
One needs the analytic form of the coefficients $B_\pm,B_3$ and
$A_\pm,A_3$ in the following identity
\begin{eqnarray}
\exp\left(A_-\hat k_-\right)\exp\left(2A_3 \hat k_3\right)\exp
\left(A_+\hat k_+\right)=\exp\left( 2B_3\hat k_3+B_+\hat k_+ +B_-\hat
k_- \right)
\label{su11}\;.
\end{eqnarray} 
By using the faithful representation of the $su(1,1)$ algebra in terms
of the Pauli matrices with $i\hat\sigma^{\pm}\equiv \hat k_\pm $,
$\hat\sigma ^3\equiv 2\hat k_3$ , Eq. (\ref{su11}) can be rewritten as follows
\begin{eqnarray}
&&\left(
\begin{array}{cc}
1 & 0 \\
iA_- & 1
\end{array}
\right)
\left(
\begin{array}{cc}
e^{A_3}& 0\\
0 & e^{-A_3}
\end{array}
\right)
\left(
\begin{array}{cc}
1 & iA_+ \\
0 & 1
\end{array}
\right)=\nonumber \\
&&\cosh \Gamma 
\left(
\begin{array}{cc}
1 & 0 \\
0 & 1
\end{array}
\right)
+\frac{\sinh \Gamma}{\Gamma}
\left(
\begin{array}{cc}
B_3 & iB_+ \\
iB_- & -B_3
\end{array}
\right)\label{pauli}\;,
\end{eqnarray}
where $\Gamma =(B_3^2-B_+B_-)^{1/2}$.  From Eq. (\ref{pauli}) one
obtains the relation
\begin{eqnarray}
B_3&=&\frac 12\frac{\Gamma }{\sinh
\Gamma}\left[\left(1+A_+A_-\right)e^{A_3} -e^{-A_3}
\right]\;,\label{b3}\\ \sinh \Gamma&=&\left\{\left
[\frac{\left(1+A_+A_-\right)e^{A_3} +e^{-A_3}
}{2}\right]^2-1\right\}^{\frac 12}\;,\label{gm} \\B_\pm&=&\frac{2A_\pm
e^{\pm A_3}}{(1-A_+A_-)e^{A_3} -e^{-A_3}}B_3 \;.\label{bpm}
\end{eqnarray}
For the purpose of the paper, we are just interested in the asymptotic
expression of the POVM ${\cal A}_g^{(\hat K)}[\d\hat H_f(gu)]$ in
Eq. (\ref{pre}) for $g\rightarrow \infty $.  By comparing
Eqs. (\ref{su11}) and (\ref{k1}) one has $A_\pm=\mp g^{-1}\lambda$ and
$A_3=-\ln (1-ig^{-1} \lambda \frac c2)$. From
Eqs. (\ref{b3}-\ref{bpm}) one obtains the asymptotic values of $B_\pm$
and $B_3$ for $g\rightarrow \infty $, namely 
\begin{eqnarray}
B_\pm\simeq \mp g^{-1}\lambda\;, \qquad B_3\simeq \frac 12
ig^{-1}\lambda c-\frac 12 g^{-2}\lambda ^2\left(1+\frac {c^2}{4}\right)\;.
\label{basy}
\end{eqnarray}
Hence, from Eq. (\ref{k1}) it follows
\begin{eqnarray}
g\d\hat H_f (gu)&\stackrel {g\gg 1} {\longrightarrow }&du\int
\frac{\d\lambda}{2\pi}\,(1+i\lambda g^{-1}\frac c4) \, e^{-i\lambda
u}\label{app}\\&\times &\exp\left\{ig^{-1}\lambda \hat K +\frac 12
\left[i\lambda g^{-1}c-g^{-2}\lambda ^2\left(1+\frac{c^2}{4}\right)\right]
 \left(a^{\dag }a+\frac
12\right)\right\}\nonumber \;.
\end{eqnarray}
By applying the amplification map to the POVM $\d\hat H_f (gu)$ through
Eqs. (\ref{uk1}) and (\ref{u2u}), one obtains ${\cal A}_g^{(\hat
K)}[\d\hat H_f(gu)]$ in Eq. (\ref{pre2}). 

\newpage
\end{document}